\documentclass[year=24,pdfa]{fmcad}

\usepackage{listings}  
\usepackage{xcolor}    
\usepackage{xspace}

\definecolor{jsonkey}{RGB}{0,128,0}   
\definecolor{jsonstring}{RGB}{163,21,21}  

\usepackage{enumitem}
\usepackage{hyperref}
\usepackage[hyphenbreaks]{breakurl}

\begin{document}
\title{Evaluating LLM-driven User-Intent Formalization for Verification-Aware Languages}

\author{Shuvendu K. Lahiri\orcid{0000-0002-4446-4777} \\
Microsoft Research, Redmond, USA\\
shuvendu@microsoft.com \\

}

\markboth{Journal of \LaTeX\ Class Files,~Vol.~14, No.~8, August~2015}%
{Shell \MakeLowercase{\textit{et al.}}: Bare Demo of IEEEtran.cls for IEEE Journals}

\maketitle

\begin{abstract}
Verification-aware programming languages such as Dafny and F* provide means to formally specify and prove properties of a program. 
Although the problem of checking an implementation against a specification can be defined mechanically, there is no algorithmic way of ensuring the correctness of the {\it user-intent formalization for programs} --- that a specification adheres to the user's intent behind the program. 
This is because intent or requirement is expressed {\it informally} in natural language and the specification is a formal artefact. 
However, the advent of large language models (LLMs) has made tremendous strides bridging the gap between informal intent and formal program implementations in the last couple of years, driven in large parts due to  benchmarks and automated metrics to evaluate different techniques.  

Recent work has developed a framework for evaluating and benchmarking the {\it user-intent formalization} problem for mainstream programming languages~\cite{endres-fse24}. However, as we argue in this paper, such an approach does not readily extend to verification-aware languages that support rich specifications (using quantifiers and ghost variables) that cannot be evaluated through dynamic execution. Previous work also required generating program mutants using LLMs to create the benchmark. We advocate an alternate, perhaps simpler  approach of {\it symbolically testing specifications} to provide an intuitive metric for evaluating the quality of specifications that can be easily instantiated with most verification-aware languages. We demonstrate that our automated metric agrees closely on a human-labeled dataset of Dafny specifications for the popular MBPP code-generation benchmark, yet demonstrates cases where the human labeling is not perfect. 
We also outline formal verification challenges that need to be addressed to apply the technique more widely. 
We believe our work provides a stepping stone to enable the establishment of a benchmark and research agenda for the problem of user-intent formalization for programs. 
\end{abstract}

\begin{IEEEkeywords}
formal verification, specifications, large language models
\end{IEEEkeywords}

\IEEEpeerreviewmaketitle

\newcommand{\ic}[1]{\begin{small}\texttt{#1}\end{small}}
\newcommand{\NLSpec}{\textit{LLM4nl2post}}
\newcommand{\GPTthree}{\ic{GPT-3.5}}
\newcommand{\GPTfour}{\ic{GPT-4}}
\newcommand{\opensource}{\ic{StarChat}}
\newcommand{\humanevalplus}{\textit{EvalPlus}}
\newcommand{\humaneval}{\textit{HumanEval}}
\newcommand{\dfj}{\textit{Defects4J}}
\newbool{longVersion}
\setbool{longVersion}{false}

\newcommand{\Comment}[1]{}
\newcommand{\shuvendu}[1]{
    {{ {\color{brown} [ \textbf{Shuvendu:} #1 ]}}}
}

 
\newcommand{\highlight}[1]{%
{\footnotesize%
\inlinebox{#1}%
}}
  
\colorlet{punct}{red!60!black}
\definecolor{background}{HTML}{EEEEEE}
\definecolor{delim}{RGB}{20,105,176}
\colorlet{numb}{magenta!60!black}

\lstdefinelanguage{json}{
    basicstyle=\normalfont\ttfamily,
    numberstyle=\scriptsize,
    stepnumber=1,
    numbersep=8pt,
    showstringspaces=false,
    breaklines=true,
    frame=lines}

\lstset{
  language=Python,
  morekeywords={}, 
  keywordstyle=\color{blue}\bfseries, 
  otherkeywords={var, method, predicate, array, seq, reads, ensures, requires, expect, assert, assume, forall, exists, &&, ||, ==>, <==>}, 
  commentstyle=\color{green}\ttfamily\itshape, 
  morecomment=[l]{//}, 
  morecomment=[s]{/*}{*/}, 
  frame=lines
}


\definecolor{dkgreen}{rgb}{0,0.6,0}
\definecolor{gray}{rgb}{0.5,0.5,0.5}
\definecolor{GrayMed}{rgb}{0.7,0.7,0.7}
\definecolor{GrayLt}{rgb}{0.9,0.9,0.9}
\definecolor{palegreen}{rgb}{0.92, 1, 0.9}
\definecolor{paleblue}{rgb}{0.92, 0.90, 1}
\definecolor{mauve}{rgb}{0.58,0,0.82}

\lstdefinestyle{PythonInline}{
  language=python,
  showstringspaces=false,
  basicstyle={\small\color{black}},
  keywordstyle=\color{blue},
  commentstyle=\color{mauve},
  stringstyle=\color{dkgreen}
}

\newcommand{\pyinline}[1]{\lstinline[style=PythonInline]!#1!}

\newcommand{\code}[1]{\pyinline{#1}}

\newcommand{\spec}[1]{\phi_{#1}}
\newcommand{\tests}[1]{\textit{T}_{#1}}
\newcommand{\btrue}{\texttt{true}}
\newcommand{\bfalse}{\texttt{false}}

\newcommand{\wrongspec}{\textsc{wrong\_spec}}
\newcommand{\weakspec}{\textsc{weak\_spec}}
\newcommand{\strongspec}{\textsc{strong\_spec}}

\section{Introduction}
\label{sec:intro}

\newcommand{\etal}{\emph{et al.}\xspace}
\newcommand{\ie}{\emph{i.e.}\xspace}
\newcommand{\eg}{\emph{e.g.}\xspace}

\interfootnotelinepenalty=10000

Formal verification is only as good as the specification it verifies.
A formal specification unambiguously defines the (possibly partial) intent behind a program, often in an declarative manner. 
Although there has been decades of research in advancing the state-of-the-art in automating the problem of verifying an implementation against a specification, relatively less attention has been spent on how to aid the generation and evaluation of specifications or formal requirements.
It is well-known that the lack of formal specifications is a significant impediment to deployment of formal verification in production code~\cite{craigen-tse95}. 
On the other hand, although a software is often accompanied by informal intent expressed in natural language comments and API documentation, these intents are seldom formally enforced on the underlying implementation. 

Large language models (LLMs) have recently demonstrated potential to bridge the gap between informal intent and formal artefact such as code~\cite{chen2021evaluating,mbpp-cite}, by performing code generation from natural language.
Such progress has been spurred in large parts through the creation of crowd-sourced benchmarks such as HumanEval~\cite{chen2021evaluating} and Mostly Basic Python Programs (MBPP)~\cite{mbpp-cite} with automated program-semantics based metrics (such as tests), unlike NLP metrics such as BLEU scores~\cite{bleu-cite}. 
For code generation, the quality of an implementation generated from informal intent is measured through the set of {\it hidden} validation tests.
We term the tests as "hidden" since they are not available to the model at the time of code generation. 
The benchmarks also provide a reference code in Python for each problem, which are also hidden from the code generation model.
These approaches rely heavily on the presence of a high-quality set of validation tests to exercise most corner cases. 
These benchmarks allow a community to evaluate their models and techniques on a common set of benchmarks automatically and measure progress. 
One can hope that the establishment of similar benchmarks and metrics for specification generation (decoupled from code generation) can enable synthesizing useful specifications from informal intent in practice. 

Motivated by such a need, Endres \etal ~\cite{endres-fse24} describe the problem of generating formal declarative specifications (namely, method postconditions) from informal intent using LLMs and automated metrics for evaluating them --- we refer to this as \textit{user-intent formalization} problem in this paper. 
For user-intent formalization, Endres \etal{} re-purpose code-generation benchmarks (such as HumanEval) and introduce two metrics (a) {\it correctness} and (b) {\it completeness}, with respect to the set of (hidden) validation tests and the (hidden) reference code.
Correctness captures that the specification satisfies the reference code for all the validation tests; completeness captures the strength of the specification to discriminate against buggy mutations of the reference code (under the set of tests).
Further, these code mutants are generated through LLMs and are grouped by the subset of tests that fail them.
They demonstrate that these automatic metrics closely resemble the quality of specifications as determined through manual analysis.
As with code generation, the approach relies on the presence of a high-quality test-suite. 

\subsection{Summary}
\label{sec:summary}
In this paper, we investigate creating benchmarks (dataset and associated automatic metrics) for user-intent formalization for {\it verification-aware languages} such as F*~\cite{fstar}, Dafny~\cite{dafny} and Verus~\cite{verus}.
A verification-aware language supports a rich program logic for expressing specifications and offers the ability to verify them statically using automated theorem provers. 
However, for most non-trivial programs, the verification requires manually decomposing the problem through the use of ghost variables, intermediate lemmas, invariants and assertions. 

We focus on the problem of evaluating specifications automatically. 
We discuss why prior approaches do not readily apply to evaluating specifications. 
Then we propose the use of automatic program verification to symbolically "test" the specifications to determine their quality. 
We have developed a prototype and applied it to a dataset of Dafny specifications from prior work~\cite{misu-fse24}.
We demonstrate that the quality of the specifications obtained through the automated means aligns well with the human-labeling of the specifications in most cases. 
Finally, we describe the unique challenges (such as quantifier instantiation) that need to be addressed to apply the technique to a larger class of problems. 

\subsection{Running example}
\label{sec:overview}
Consider a snippet of the JSON specification of an example from the MBPP dataset for code generation from natural language in Python~\cite{chen2021evaluating}.

\begin{lstlisting}[language=json, basicstyle=\ttfamily\scriptsize]
"prompt": "Write a function to find the shared elements from the given two lists.",
"code": "def similar_elements(test_tup1, test_tup2):\n  ....",
"test_list": [
    "assert set(similar_elements((3, 4, 5, 6),(5, 7, 4, 10))) == set((4, 5))",
    ...
]
\end{lstlisting}
The dataset comes with a set of problems, each containing a natural language prompt \code{prompt}, the reference code \code{code} as well a set of 3 tests in \code{test\_list}.
Misu \etal{}~\cite{misu-fse24} port these requirements to Dafny (call it MBPP-DFY), including slight change to the prompt, explicitly representing the Dafny method signature (with a slightly changed name), as well as the test cases~\cite{mbpp-san-dfy-228-all-task-test.json}.
\footnote{We take the snapshot of the repository at COMMIT a57ce24.} 
\begin{lstlisting}[language=json, basicstyle=\ttfamily\scriptsize]
"task_description": "Write a method in Dafny to find the shared elements from the given two array.",
"method_signature": "method similarElements (arr1:array<int>, arr2:array<int>) returns (res: array<int>)",
"test_cases": {
    "test_1": "var a1:= new int[] [3, 4, 5, 6];\nvar a2:= new int[] [5, 7, 4, 10];\nvar e1:= new int[] [4, 5];\nvar res1:=similarElements(a1,a2);\nassert arrayEquals(res1,e1);",
      ...
}
\end{lstlisting}

Misu \etal{} use GPT-4~\cite{gpt4-paper} and other language models (with sophisticated prompting) to generate a pair of Dafny specification and implementation, and retain generations where the generated specification is provable for the generated implementation. 
Below we show the specification component of the GPT-4 generated Dafny artefact for this example (with slightly altered method name and signature over JSON)~\cite{task-id-2.dfy}.
The two postconditions (marked with \code{ensures}) define the specification and use an auxiliary predicate \code{InArray}.
\begin{lstlisting}[language=Python, basicstyle=\ttfamily\scriptsize]
predicate InArray(a: array<int>, x: int)
reads a
{ exists i :: 0 <= i < a.Length && a[i] == x }

method SharedElements(a: array<int>, b: array<int>) 
               returns (result: seq<int>)
  ensures forall x :: x in result ==> 
                      (InArray(a, x) && InArray(b, x))
  ensures forall i, j :: 0 <= i < j < |result| ==> 
                      result[i] != result[j]
{
   ....Dafny implementation...
}
\end{lstlisting}

Although the implementation satisfies the specification, there is no evidence that the specification indeed captures what the user intended (and specified implicitly in the hidden test cases). 
The authors perform {\it manual} review of the specifications and mark them as either $\{\wrongspec, \weakspec, \strongspec\}$ to denote if the specification is respectively, inconsistent, weakly or strongly consistent with the intent expressed in the natural language and tests~\cite{gpt4-generations-mbpp-dfy}. 
Therefore, although this work addresses generating verified Dafny programs from informal intent, it relies completely on a human to ensure that the specification matches the intent.
In particular, the framework allows for the Dafny code to only satisfy a vacuous specification (such as \code{ensures}  \btrue) or worse,  an incorrect one.
This aspect renders this dataset unsuitable as an automated benchmark for specification generation in verification-aware languages.
In other words, if a language model generates a specification that is non-equivalent to the ground-truth specification labeled by the user, we cannot determine if it is incorrect without requiring a user. 
This also makes the evaluation subjective. 
Since verification of (generated) Dafny code is only as good as the specification, this makes the dataset unsuitable for an automated benchmark for verified code generation as well. 


\subsection{Proposal: Symbolically testing specifications}
Our objective in this work is to define  metrics that can be automatically evaluated to determine the quality of a specification. 
At the same time, we would also like to establish that these metrics correspond well to what a manual labeling would establish. 
Further, we decouple the problem of specification generation from  code generation, to be able to (a) use the specifications to find bugs in underlying or generated code~\cite{endres-fse24}, or (b) use the validated specifications to refine the prompts for code generation.

We first argue that the none of the prior approaches are readily applicable to yield an automatic metric:
$(i)$ First, running the tests on the implementation without specifications will not rule out vacuous specifications.
$(ii)$ Second, one cannot apply the approach of Endres \etal{}  for rich specifications that contain ghost state or constructs that cannot be evaluated at runtime (such as a universal/existential quantifier). 
$(iii)$ One can ignore tests and attempt to statically verify the synthesized specification directly against the hidden reference code. This can still allow vacuous specifications such as \btrue. Further, it is unlikely that such verification would  be automatic for non-trivial specifications given the need to infer intermediate lemmas and invariants.
$(iv)$ One can instead provide a hidden reference specification and check the candidate specification for  semantic equivalence. 
However, such a metric would be too strict as it would not be able to distinguish weak and vacuous specifications (such as $\btrue$) from strong (yet incomplete) specifications. 

Next, we propose a method for evaluating user-intent formalization for verification-aware languages that leverages the validation tests and symbolic verification capabilities of these languages.
Given a set of tests consisting of a set of {\it consistent} input-output pairs, we define the correctness and completeness metrics purely over the tests, without the need for the reference code.
Of course, just as in the case of code generation, the technique for determining specification quality assumes a high-quality set of validation tests. 

Consider a method with signature \code{m(x):y} denoting the name \code{m}, input parameters \code{x} and output parameters \code{y}, a candidate postcondition/summary specification $\spec{}$(\code{x}, \code{y}) and a set of input-output tests $\tests{}$.

\subsubsection{Correctness}
A postcondition $\spec{}$(\code{x}, \code{y}) is correct (or sound) with respect to $\tests{}$ if it is consistent with all the input-output pairs in $\tests{}$.
In other words, for each $(i, o) \in \tests{}$ the following Hoare-triple~\cite{hoare-logic-cite} holds.
$$
\models \{\btrue\} \ \code{x} := i; \ \code{y} := o; \ \{\spec{}(\code{x}, \code{y})\}
$$

\subsubsection{Completeness}
The completeness measure for a specification $\spec{}$ given   $\tests{}$ is the fraction of output mutations of the tests in $\tests{}$ that $\spec{}$ is inconsistent with.
Let $\tests{1} \doteq \bigcup_{(i,o) \in \tests{}} \bigcup_{o' \neq o} \{(i,o')\}$ be a finite set of mutations of $\tests{}$ that mutate the output values for the given inputs.
In this work, we restrict the set of mutants per input $i$ to a fixed number (5 for this paper).
Let $\tests{2} \subseteq \tests{1}$ be the largest subset such that for each $(i,o') \in \tests{2}$, the following Hoare-triple {\it does not} hold:
$$
\not\models \{\btrue\} \ \code{x} := i; \ \code{y} := o'; \ \{\spec{}(\code{x}, \code{y})\}
$$
Then the completeness measure of $\spec{}$ with respect to $\tests{}$ is $|\tests{2}|/|\tests{1}|$ (this is inspired by kill-set in {\it mutation-testing} literature~\cite{jia2010analysis}).
For interested readers, we also contrast these with a plausible proposal in Appendix~\ref{sec:alternate-hoare}.

Let us demonstrate an implementation of these Hoare triples as Dafny programs for our running example. 
Consider the first test (see JSON input in Section~\ref{sec:overview}) that asserts that \code{SharedElements} should return the set $\{4, 5\}$ for input arrays $[3, 4, 5, 6]$ and $[5, 7, 4, 10]$.
For correctness against this test, we create the following Dafny program by providing a definition of \code{SharedElements} that performs the soundness check:
\begin{lstlisting}[language=Python, basicstyle=\ttfamily\scriptsize]
predicate InArray(a: array<int>, x: int)
reads a
{  exists i :: 0 <= i < a.Length && a[i] == x }

method SharedElements(a: array<int>, b: array<int>) 
               returns (result: seq<int>)
  ensures forall x :: x in result ==> 
                      (InArray(a, x) && InArray(b, x))
  ensures forall i, j :: 0 <= i < j < |result| ==> 
                      result[i] != result[j]
{
  var a1 := new int[] [3, 4, 5, 6];
  var a2 := new int[] [5, 7, 4, 10];
  assume {:axiom} a[..a.Length] == a1[..a1.Length];
  assert a[0] == a2[0] && .... && a[3] == a2[3];
  assume {:axiom} b[.. b.Length] == a2[..a2.Length];
  assert b[0] == a2[0] && .... && b[3] == a2[3];
  result := [4, 5];
}
\end{lstlisting}

The Dafny program above is identical to the program in Section~\ref{sec:overview}, except for the body of \code{SharedElements} specified within the curly braces $\{\ldots\}$.
Instead of the original implementation of the method, we model the Hoare-triple for correctness described above. 
The precondition of the Hoare-triple ($\btrue$) translates to \code{assume true} which is dropped. 
The input assignment $\code{x} := i;$ is modeled as assignments of the input $i$ to temporary variables \code{a1, a2}, followed by constraining the actual parameters \code{a, b} respectively. 
For Dafny, this amounts to saying that two arrays \code{a} (respectively, \code{b}) and \code{a1} (respectively, \code{a2}) are equal on all elements up to their lengths (which implies that the lengths are identical as well).
The redundant asserts are needed for the verifier to trigger the  quantifiers used in \code{InArray} to enable the proof.
Finally, the output parameter \code{result} is assigned one of the expected values (\code{[5,4]} is also an acceptable value). 
The Dafny program symbolically checks (using Satisfiability Modulo Theories solvers~\cite{smt-lib}) that the specification (provided by the \code{ensures} statements) holds for the specific input and output.

However, the above verification only proves that the specification is {\it correct} for the test; a vacuous specification \code{ensures true} would also be verified.
For completeness, we mutate the output value in \code{result} in several ways and check that the verification fails. 
In this case, mutating the value in \code{result} to \code{[6]} would fail the first postcondition, since \code{6} is not present in both the input arrays. 

For this program, the above specification is marked as $\strongspec$ by the authors of MBPP-DFY.
However, our automated test harness coupled with mutation discovers that the specification is \textit{incomplete} (score 0.6).
When we check the above program with either of the two mutations  \code{result := [4]} or \code{result := [5]}, the specification still verifies.
This is because the implication \code{==>} in the first postcondition only checks that values in \code{result} is present in both the arrays; it does not check that all such common values are present in \code{result}!
Thus our automated metric is able to assign the above specification a lower score than the full functional specification for this example where the \code{==>} is replaced by \code{<==>}.

\paragraph*{Note} One may caution against using a verification failure (as used for the completeness checks) as a means to show that the corresponding Hoare-triple does not hold. In other words, given that deductive verifiers are sound, but not necessarily complete, a verification failure is typically interpreted as an unknown outcome. However, note that we check for completeness for a specification $\spec{}$ against an input-output example $(i, o')$ only if  $\spec{}$ is \textit{proved correct} for all the tests in $\tests{}$ (including the original test $(i,o)$). Since the two verification conditions differ only in a concrete value (between $o$ and $o'$), we conjecture that the underlying reasoning (including quantifiers instantiated) would be quite identical in most cases, and the verification failure strongly indicates that the Hoare-triple does not hold.

\section{Implementation and evaluation}
\label{sec:results}

We report ongoing work in implementing the two metrics for the MBPP-DFY dataset mentioned earlier in Section~\ref{sec:overview}.
Our tool (a 400 line Python script) consumes the method signature, test cases and the candidate specification from the JSON and Dafny files, and creates Dafny programs for verifying the correctness and completeness for each test.
It then invokes the Dafny verifier, and reports the aggregate correctness as well completeness score for each specification, averaged over the different test cases; it finally compares the metrics against the labels provided by authors.
Of the 153 problems with specifications (a subset written by humans), we have  managed to apply our tool to check 64 of the specifications (at the time of writing).
In other words, for these 64 examples, our tool is able to verify the soundness of the specifications over the set of validation tests. 
We have released the scripts, dataset and outputs at the website \url{https://github.com/microsoft/nl-2-postcond}.

\subsection{Mutating values} 
Each of these examples contain 3 test cases in Dafny format, and we consider up to 5 distinct mutants of the output values for each test. 
We currently have a simple mutation scheme for the output values. 
For Booleans,  we flip the value between $\btrue$ and $\bfalse$.
For integers, we choose a random value between 1 and 10, and randomly add or subtract it.
For strings, we choose a random character and either replace one of the characters or append it to the string.
For arrays (we only restrict to integer arrays and sequences), we choose between dropping an element or inserting a random value at a randomly chosen index.
For our running example of \code{SharedElements}, this allows us to create a mutant test case with the array value $[4]$ that demonstrates that the GPT-4 generated specification (marked as $\strongspec$) is not the most precise specification for this problem. 

\subsection{Results}
We briefly report some details of the evaluation, and outline the challenges to handle the remaining examples. 

We find that for large majority of examples, the manual labels align with the metrics computed by our tool. 
That is, for the incorrect specifications, our tool reports a verification failure for correctness check, and report a high completeness score (usually $> 0.66$) for precise specifications.
In other words, barring a few exceptions below, all the specifications that are marked \strongspec{} have a completeness score above $0.66$, and all the \weakspec{} have a score below $0.66$.

This provides evidence that the automated metrics serve as a good proxy for the user label. 
One such incorrect specification generated by GPT-4 (and correctly labeled) is the problem of \code{RemoveDuplicates} ("task\_id" 572) (also the motivating example by Endres \etal{}~\cite{endres-fse24} demonstrating ambiguity of natural language) where the task is to remove all elements with duplicates, but GPT-4 interprets it as the problem of retaining a single copy of each value. 
Similarly, for \code{countSubstrings} ("task\_id" 61), that counts the number of substrings whose sum equals their length, our completeness check gives a low score to the specification (correctly labeled $\weakspec$)  that only ensures \code{count} is a non-negative number. 

In addition to the \code{SharedElements} ("task\_id" 2), we found at least 2 more cases where the specification (labeled $\strongspec$) is weaker than the most precise one. 
These include the cases for \code{maxAbsDiff} ("task\_id" 145), and \code{removeElements} ("task\_id" 161). 
\code{maxAbsDiff} is expected to compute the maximum difference between any two elements in the list; the provided specification only ensures that the result upper bounds the difference between any two elements, but is not the precise bound. 
\code{removeElements} is expected to remove the elements in one array from another; the specification only ensures that any element in the result array is present in the first array and not the second but fails to ensure that the result array is the precise difference (can be ensured by changing \code{==>} by \code{<==>}). 

Interestingly, we also discovered a few cases where the correctness check failed for specifications labeled $\strongspec$. 
These include examples with "task\_id"s 234, 240, and 445. 
For these examples, the authors had accidentally introduced bugs while copying from the test cases in Python. 
For instance, for "task\_id" 234 that computes the cube of a number, the authors mistakenly copied the value 25 instead of 125 for the cube of 5. 
We noticed that a couple of these errors have been fixed in the latest commit at the time of writing (fb4f53e), but a few still remain (e.g., 445 where the value 32 is replaced by 31). 

Our preliminary results demonstrates that the automatic metrics not only helps to add objectivity to the manual labeling, argues for having a quantitative metric for completeness (instead of a Boolean $\weakspec$ vs $\strongspec$), but also clearly demonstrates the potential to serve as a metric for evaluating specifications for a benchmark. 

\subsection{Limitations}
Our implementation currently has a few limitations that precludes analyzing all the 153 specifications correctly.
A large class of them stem from inaccurate parsing through a simple regex based parsing of the Dafny files. 
We expect these (e.g., handling of 2D arrays) to be addressed as we improve the parsing through Dafny AST-based analysis. 
However, a few fundamental challenges (related to failure to perform proofs of correct specifications) remain that requires further investigation. 
We note a couple of them here:
\paragraph{\it Recursive predicates.} Specifications for tasks such as 105 (counting number of true Booleans in an array) require the use of recursive functions \code{countTo} that expresses the count of an array in terms of \code{countTo} of the tail of the array. 
Dafny uses a notion of "fuel" to control the level of unfolding of such predicates, which suffices for inductive proofs.
However, for concrete arrays in test cases, these recursive functions need to be unrolled proportional to the length of the array. 
\paragraph{\it Quantifier instantiation.} Specifications for tasks such such as 3 (that classifies a number as a non-prime) uses existential quantifiers.
For example, checking a number $n$ as non-prime requires checking for the existence of a smaller number $k$ that divides $n$. 
Unlike their usage in inductive proofs where a loop establishes the inductive hypothesis to only instantiate the quantifier on a few expressions, testing a symbolic specification against a concrete value is sometimes non-trivial.
For example, to show that the value 97 is a prime, the quantifier needs to be instantiated on all values up to 96. 

We have already automated a few failed proofs due to quantifier instantiation by adding intermediate redundant assertions equating two arrays on every index (see \code{assert} statements in the body of \code{SharedElements} in Sec~\ref{sec:overview}).  
In future work, we plan to automate the inference of fuel amount and quantifier instantiation witnesses when ranging over the indices of an array. 
\section{Related work}
\label{sec:related}

Although the literature on specification mining is fairly rich, it has historically focused on techniques that infer specifications of an implementation through  dynamic~\cite{ernst1999dynamically} and recently neural techniques~\cite{pei2023can}. 
The objective of these techniques is to learn invariants from a few runs that generalize to unseen executions.
Static~\cite{houdini,lahiri-inference-cav09,padon-popl22} and more recently neural approaches~\cite{code2inv-cav20,kamath2023finding} have also been applied to the problem (inductive) invariant generation to aid program proofs. 
Finally, recent works have explored the generation of programs, specifications and proofs from natural languages~\cite{misu-fse24,sun2024cloverclosedloopverifiablecode,loughridge2024dafnybenchbenchmarkformalsoftware}.
On the other hand, work on user-intent formalization (i.e., translating natural language comments to specifications)~\cite{blasi2018translating,endres-fse24} synthesize the intended specification, even in the absence of an implementation. 
The problem of generating {\it test oracle} assertions have also been investigated~\cite{dinella2022toga}, but they only apply to single input or test prefix.
TiCoder~\cite{lahiri2022interactive} introduced the term "user-intent formalization" over test cases for interactive code generation, and alluded to metrics for measuring quality of weak specifications (namely, tests) in terms of  correctness (user acceptance) and completeness (prunes buggy codes).
The work on {\it autoformalization}~\cite{wu2022autoformalization} for translating natural language comments to mathematical theorems is closest to our work.
However, these techniques do not use semantic checks (such as tests and symbolic verification) to ensure the quality of the generated formal mathematical statements; instead they leverage textual similarity metrics such as BLEU.

\section{Conclusion}
\label{sec:concl}
In this paper, we motivate the problem of evaluating user-intent-formalization for verification-aware languages. 
We demonstrate that the idea of symbolically testing specifications against validation tests can provide an automated metric for a benchmark.
We plan to curate a benchmark with a large fraction of examples from MBPP-DFY dataset using the above metric. 
We hope such benchmarks will accelerate the research on specification generation from informal intent in verification-aware languages.
This when coupled with work on program and proof synthesis (given a specification) can greatly lower the cost required to create formally verified  modules in the future~\cite{popaiarxiv24}.

\appendices

\section{Extended version of the JSON examples}
\begin{lstlisting}[language=json, basicstyle=\ttfamily\scriptsize, keywordstyle=\color{jsonkey}, stringstyle=\color{jsonstring}]
"prompt": "Write a function to find the shared elements from the given two lists.",
"code": "def similar_elements(test_tup1, test_tup2):\n  ....",
"test_list": [
    "assert set(similar_elements((3, 4, 5, 6),(5, 7, 4, 10))) == set((4, 5))",
    "assert set(similar_elements((1, 2, 3, 4),(5, 4, 3, 7))) == set((3, 4))",
    "assert set(similar_elements((11, 12, 14, 13),(17, 15, 14, 13))) == set((13, 14))"
]
\end{lstlisting}

\begin{lstlisting}[language=json, basicstyle=\ttfamily\scriptsize, keywordstyle=\color{jsonkey}, stringstyle=\color{jsonstring}]
"task_description": "Write a method in Dafny to find the shared elements from the given two array.",
"method_signature": "method similarElements (arr1:array<int>, arr2:array<int>) returns (res: array<int>)",
   "test_cases": {
      "test_1": "var a1:= new int[] [3, 4, 5, 6];\nvar a2:= new int[] [5, 7, 4, 10];\nvar e1:= new int[] [4, 5];\nvar res1:=similarElements(a1,a2);\nassert arrayEquals(res1,e1);",
      "test_2": "var a3:= new int[] [1, 2, 3, 4];\nvar a4:= new int[] [5, 4, 3, 7];\nvar e2:= new int[] [3, 4];\nvar res2:=similarElements(a3,a4);\nassert arrayEquals(res2,e2);",
      "test_3": "var a5:= new int[] [11, 12, 14, 13];\nvar a6:= new int[] [17, 15, 14, 13];\nvar e3:= new int[] [13, 14];\nvar res3:=similarElements(a5,a6);\nassert arrayEquals(res3,e3);"
    }
\end{lstlisting}

\section{Alternate proposal}
\label{sec:alternate-hoare}
A reader may wonder if the following check achieves a similar objective for checking completeness:
$$
\models \{\code{x} == i \wedge \spec{}(\code{x}, \code{y})\} \ \code{skip} \ \{\code{y} == o\}
$$
There are two issues with this formulation: (a) this only provides a Boolean metric that only rewards a precise specification that constrains \code{y} to a unique value $o$. For example, it would award the strong (yet imprecise) specifications of \code{SharedElements} to 0. Secondly, (b) this formulation is also too strong when the precise functional specification allows for non-determinism in the output. Consider the running example of \code{SharedElements}, where the specification allows for the output to be one of either $\{[4, 5], [5, 4]\}$ for the given inputs.


\ifCLASSOPTIONcaptionsoff
  \newpage
\fi

\bibliographystyle{abbrv}
\bibliography{main}

%

\end{document}